# $\gamma p \to V p$ and $\gamma^* p \to V p$

P.V.Landshoff

DAMTP, University of Cambridge

Exclusive vector meson production from real and virtual photons is a good probe of pomeron physics. In particular, it sheds light on how the soft pomeron is related to nonperturbative QCD, and how it couples to heavy flavours. It also raises the question whether there are two pomerons, a soft one and a hard one.

*The soft pomeron*

All total cross-sections are found to rise at high energy at the same rate[1], namely like $s^{0.08}$. This is seen in figure 1. In addition to the rising term, one needs a term that falls like approximately $s^{-\frac{1}{2}}$, which is well understood to result from $\rho, \omega, f_2, a_2$ exchange. The $s^{0.08}$ term is said to be the result of soft pomeron exchange, and one of the aims of current research is to understand the soft pomeron in terms of nonperturbative QCD. It is possible that there is also a hard pomeron[2]; if this exists, it is decribed by perturbative QCD.

According to the fits shown in figure 1, the ratio of the strengths of the pomeron-exchange term in $\pi p$ and $pp$ scattering is

$$\frac{13.6}{21.7} \approx \frac{2}{3}$$

This is an indication that pomeron exchange obeys the *additive quark rule*: the pomeron couples to the separate valence quarks in a hadron. The strength in $Kp$ scattering is a little less than in $\pi p$; one possible explanation is that the pomeron couples more weakly to heavier quarks than to light quarks, 25% weaker in the case of the $s$ quark.

*$\gamma p \to V p$*

A calculation of diffractive $\rho$ photoproduction, with essentially no free parameters, can be obtained by assuming vector meson dominance and the additive quark model. The simplest version of vector meson dominance tells us that the forward cross section for $\gamma p \to \rho^0 p$ is a constant times that for $\rho^0 p \to \rho^0 p$. The constant is found from the $e^+ e^-$ decay width of the $\rho$. Even if this simplest version of VMD is refined to include also the contribution from off-diagonal terms, for example $\rho' \to \rho$, vector dominance is not an exact science. So I use the simplest version. The additive quark model then equates the $\rho^0 p \to \rho^0 p$ amplitude to that for $\pi^0 p \to \pi^0 p$. To avoid needing to make any assumptions about the slope of the differential cross section, so effectively assuming knowledge of the $\rho$ form factor, it is more convenient in the first instance to calculate the forward differential cross section rather than the total cross section. The energy dependence of the naive prediction I have outlined is correct, but the normalisation is too high. Multiplying by a factor of 0.84 provides an excellent description of all the data[3][4]. I show this in figure 2.

This normalisation difference may be explained by finite-width corrections to the $\rho \to e^+ e^-$ decay rate[5]: because of the photon propagator the mass spectrum of the $\rho$'s produced in $e^+ e^-$ annihilation is distorted towards the low-mass side of the $\rho$ peak.

---

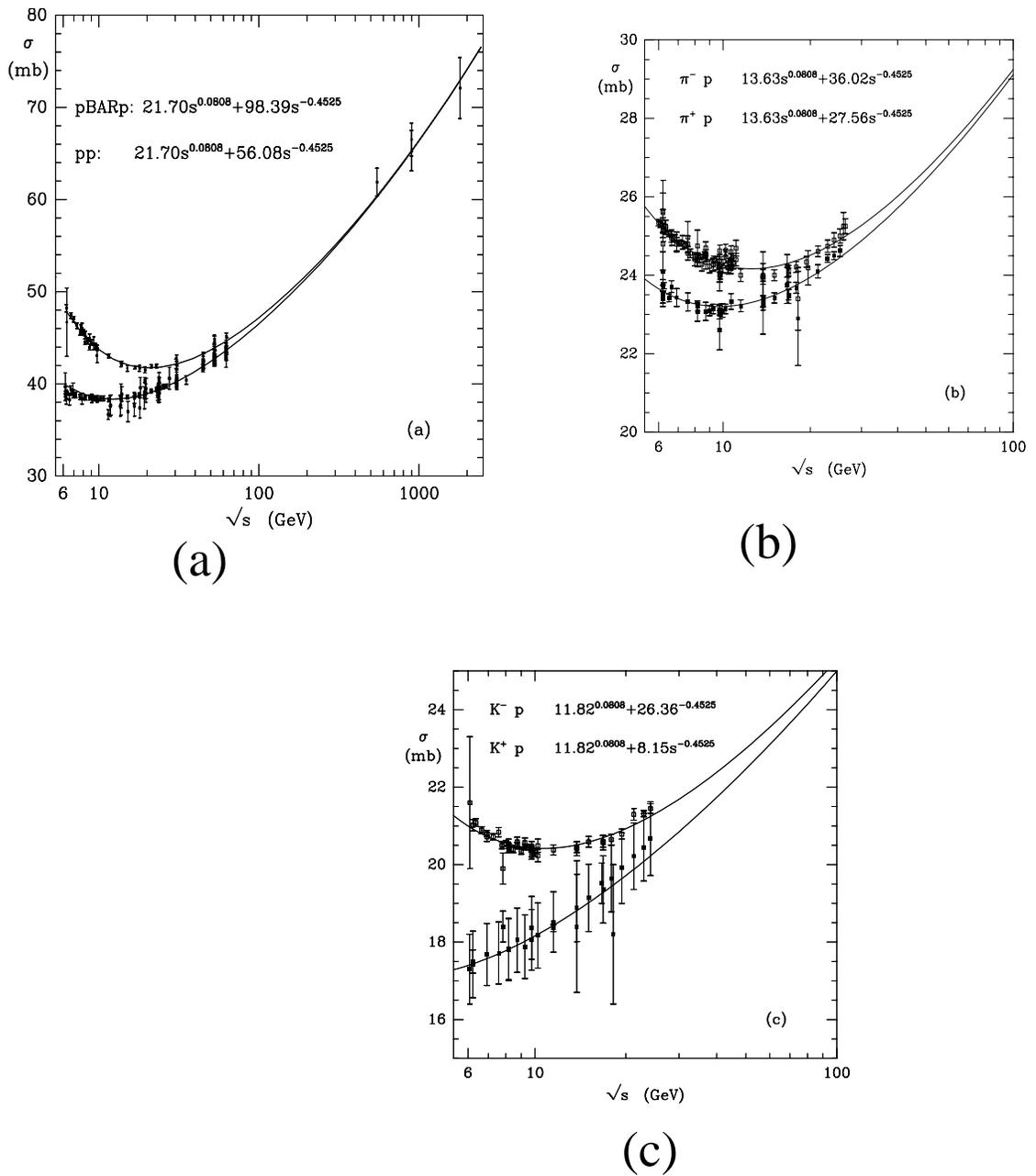

Figure 1: Total cross-sections.

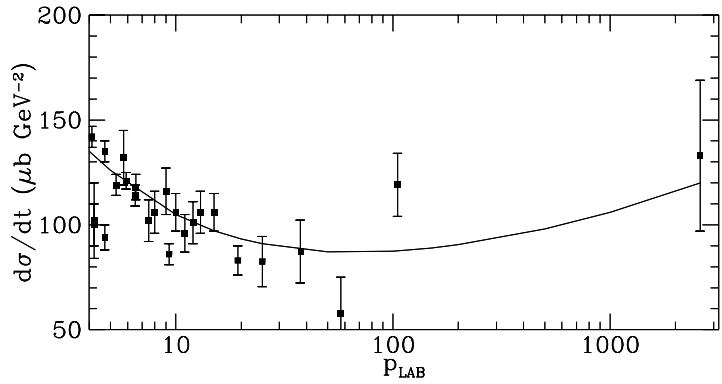

Figure 2: Data for the differential cross section for $\gamma p \to \rho p$ at $t = 0$ with soft-pomeron-exchange fit[1]

In the case of $\gamma p \to \phi p$, a corresponding calculation gives a result that turns out to be too large[3][4]. One may reduce it by a factor of almost 2 by using what I have said above about the relative weakness of the coupling of the pomeron to the strange quark (we need the square of the amplitude, so the suppresion factor is $(\frac{3}{4})^2$). But this is still too large: we need a further factor 2 suppression. Figure 3 shows a comparison with the data when we include this factor. The simplest way to explain it is to say that vector dominance for the $\phi$ is just not correct. After all, we have no theoretical understanding of why vector dominance should ever be correct, and assuming it for the $\phi$ requires the belief that the very simplest extrapolation can be made over a very large distance in $q^2$.

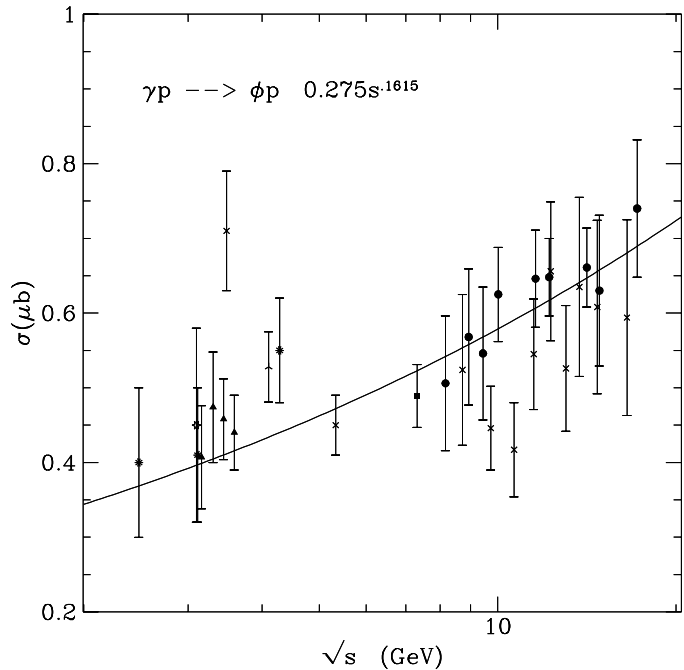

Figure 3: Data for $\gamma p \to \phi p$ with soft-pomeron-exchange fit[1]

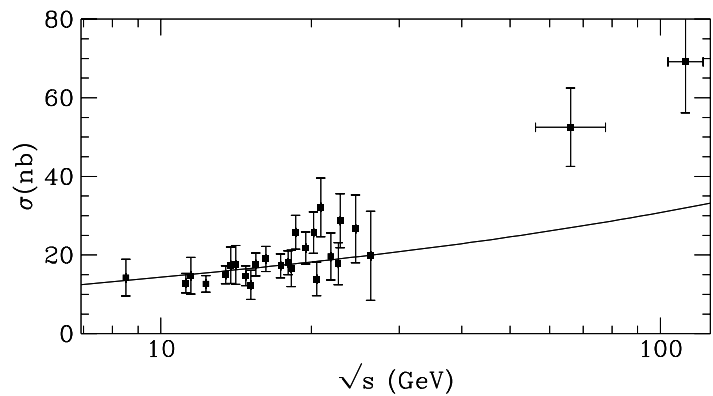

Figure 4: Data for $\gamma p \to J/\psi p$ with soft-pomeron-exchange fit[6]

If vector dominance is no good for the $\phi$, it will surely be even less good for the $J/\psi$. As I will explain, there is good reason to think that is the case. Figure 4 shows a soft-pomeron-exchange fit[6] to the data for $\gamma p \to J/\psi \, p$. If $\psi$-dominance is to be believed, we can deduce the strength of the pomeron coupling to the $J/\psi$ and so conclude that

$$\sigma^{TOT}(J/\psi \, p) \approx \frac{1}{10}\sigma^{TOT}(\rho p) \tag{1}$$

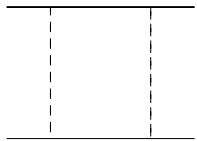

Figure 5: Exchange of nonperturbative gluons between quarks

*Theory of the soft pomeron*

The crudest model for soft pomeron exchange is the simple exchange of a pair of gluons between quarks (figure 5). In order that this model may well approximate the observed properties of the soft pomeron, the $k^2 = 0$ pole in the perturbative gluon propagator $D(k)$ must have been removed by nonperturbative confinement effects, so that the integral

$$\int_{-\infty}^{0} dk^2 \, [D(k^2)]^2 \tag{2}$$

converges. Then one finds[7] that, in this model, the pomeron couples to single quarks like an isoscalar $C = +$ photon, as its phenomenology suggests is the case[8], and also one can understand why it can obey the additive-quark rule. However, the model is too crude to reproduce the required energy dependence $s^\epsilon$ of the total cross-section: one has to put this factor in by hand. In due course, one might hope to calculate this factor by refining the model so as to include $t$-channel iterations of the simple graph, and more complicated effects, but so far this has not been possible[9].

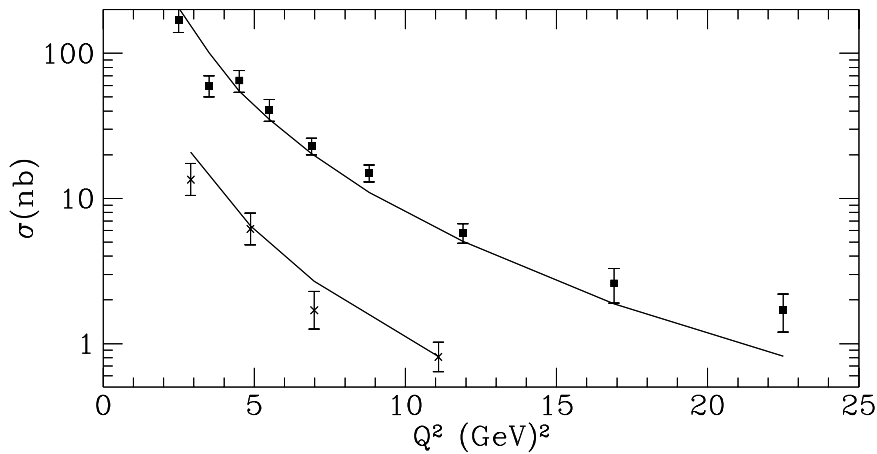

Figure 6: NMC data[10] for $\gamma^* p \to \rho p$ and $\gamma^* p \to \phi p$ with predictions[11]

$\gamma^* p \to V p$

Figure 6 shows NMC data[10] for the exclusive processsses $\gamma^* p \to \rho p$ and $\gamma^* p \to \phi p$. The curves are the predictions[11] from the simple two-gluon-exchange model for pomeron exchange. They are absolute predictions, in the sense that the only free parameter is the contribution from small values of $k^2$ to the ratio

$$\frac{\int_{-\infty}^{0} dk^2 \, 2k^2 [D(k^2)]^2}{\int_{-\infty}^{0} dk^2 \, [D(k^2)]^2} \tag{3}$$

Because the small-$k^2$ behaviour of $D(k^2)$ is controlled by nonperturbative effects, and the scale of nonperturbative effects is typically 1 GeV$^2$, we expect the ratio to be close to this value, and this was used to calculate the curves.

The curve for $\gamma^* p \to \phi p$ includes the suppression factor $(\frac{1}{4})^2$ which I have said characterises the coupling of the pomeron to the $s$ quark. If we calculate $\gamma^* p \to J/\psi\, p$ in the same model, and include no suppression factor, we find that it is larger than $\gamma^* p \to \rho p$ at high $Q^2$. Some old EMC data[12] seem to support this, but EMC made no attempt to remove from these data the contamination from events where the proton has broken up. NMC found[10], by a careful study of $\gamma^* p \to \rho p$, that this contamination is a major experimental problem. It is likely, therefore, that $\gamma^* p \to J/\psi\, p$ is smaller than $\gamma^* p \to \rho p$, so that one does need a suppression factor in the coupling of the pomeron to the $c$ quark. However, one certainly does not want a suppression as much as $(\frac{1}{10})^2$, which is the factor one would deduce from (1). This is why I believe that $\sigma^{TOT}(J/\psi\, p)$ is somewhat larger than is given in (1), and that vector dominance for the $J/\psi$ is not valid[3].

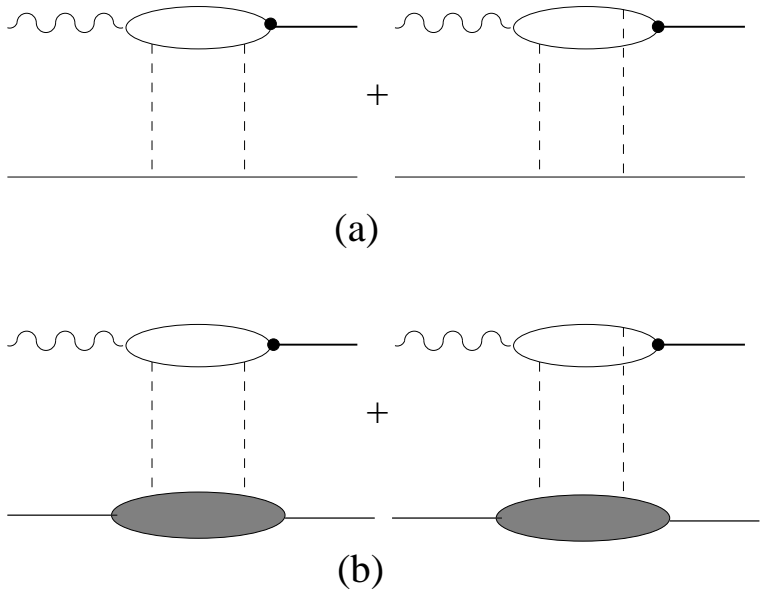

Figure 7: (a) simple model for $\gamma^* p \to \rho p$ and (b) refinement where the simple lower structure is replaced with the complete gluon structure function of the proton

*Gluon structure function*

The curves in figure 6 correspond[11] to the sum of the two diagrams in figure 7a. In this simple model, the energy dependence $W^{2\epsilon}$ must be inserted by hand. If $\epsilon = 0.08$, it predicts that the cross-section will approximately double between $W = 10$ GeV and 100 GeV. If a faster rise than this is found, one would conclude either that the soft pomeron becomes steadily less soft as $Q^2$ increases[13], that is $\epsilon$ grows steadily with $Q^2$, or that $\epsilon$ remains fixed at 0.08 but there is another term that must be added to the soft-pomeron one, with a larger power of $W$ and with a strength that becomes relatively more important as $Q^2$ increases. This might be the BFKL pomeron[2]. In either case, one would expect that the usual evolution of the forward slope $b$ of $d\sigma/dt$ will have been affected, and that we no longer have[8]

$$\frac{db}{d(\log s)} = 2\alpha'$$

$$\alpha' = 0.25 \text{ GeV}^{-2} \qquad (4)$$

The diagrams of figure 7a couple the gluons to the quark in the simplest way, as in figure 5. A more refined model is to introduce[14] the complete $ggqq$ amplitude, as in the diagrams of figure 7b. Then one no longer has to put in the energy dependence $W^{2\epsilon}$ by hand, it is contained in this amplitude. In figure 7a the lower line is a quark, which is one of the valence constituents of the target proton, but in figure 7b we may take the lower line to be the proton itself. Then the lower amplitude is almost the gluon structure function. This would seem to offer an attractive means to measure the gluon structure function, because its square would come into the differential cross-section for $\gamma^*p \to Vp$. Of course, the gluon structure function is the imaginary part of the lower amplitude evaluated at zero momentum transfer (because it is defined through the optical theorem). It is easy to correct for the fact that we need also the real part:

$$\mathrm{Re}T = \mathrm{Im}T \; \tan(\tfrac{1}{2}\pi\epsilon) \qquad (5)$$

But correcting for the fact that in $\gamma^*p \to Vp$ the momentum transfer is nonzero is likely to be more of a problem[3]. It is true that, at high energy, $t_{min}$ in $\gamma^*p \to Vp$ is very small, so that it is only a short extrapolation to $t=0$, but unfortunately this takes one to lightlike momentum transfer $\Delta$. That is, although $t=\Delta^2 \to 0$, the vector $\Delta$ is far from 0. Indeed, simple kinematics show that $q.\Delta \approx \tfrac{1}{2}Q^2$. This reflects itself in the fact that the momenta of the two gluon lines are not equal, as they would be for the gluon structure function, and for neither of them is the component in the direction of the nucleon momentum equal to $xp$; rather their *difference* $\Delta$ has component $xp$. Thus, even if the gluon structure function is involved, we do not know what $x$ value to use for it. The conclusion is that, at best, the connection between $\gamma^*p \to Vp$ and the gluon structure function is only approximate[3].

This research is supported in part by the EU Programme "Human Capital and Mobility", Network "Physics at High Energy Colliders", contract CHRX-CT93-0357 (DG 12 COMA), and by PPARC

## References

1. A Donnachie and P V Landshoff, Physics Letters B296 (1992) 227

2. E A Kuraev, L N Lipatov and V S Fadin, Sov Physics JETP 44 (1976) 443
   L V Gribov, E M Levin and M G Ryskin, Physics Reports 100 (1983) 1

3. A Donnachie and P V Landshoff, Physics Letters B348 (1995) 213

4. G A Schuler and T Sjostrand, Physics Letters B300 (1993) 169

5. G Gounaris and J J Sakurai, Physical Review Letters 21 (1968) 244
   F M Renard, Nuclear Physics B15 (1970) 267

6. M Diehl, private communication

7. P V Landshoff and O Nachtmann, Z Physik C35 (1987) 405

8. P V Landshoff, *The two pomerons*, HEP-PH 9410250

9. D A Ross, J Phys G15 (1989) 1175

10. NMC collaboration: P.Amaudruz et al, Z Physik C54 (1992) 239

11. A Donnachie and P V Landshoff, Physics Letters B185 (1987) 403 and Nuclear Physics B311 (1989) 509
    J R Cudell, Nuclear Physics B 336 (1990) 1

12. EMC collaboration:J.J.Aubert et al, Physics Letters B161 (1985) 203;
    J.Ashman et al, Z Physik C39 (1988) 169

13. A Capella, A Kaidalov, C Merino and J Tran Thanh Van, HEP-PH 9407372

14. S Brodsky et al, Physical Review D50 (1994) 3134